\documentclass[t11pt,reprint,preprint,preprintnumbers]{revtex4}
\usepackage{pst-node}
\usepackage{auto-pst-pdf}
\usepackage{epstopdf}
\usepackage{graphicx}
\usepackage{amsmath}
\usepackage{amssymb}
\usepackage{xcolor}
\usepackage{xfrac}

\newcommand{\bra}[1]{\left\langle #1 \right|}
\newcommand{\ket}[1]{\left|#1\right\rangle}

\newcommand{\brake}[1]{\left\langle#1\right\rangle}

\DeclareMathOperator*{\maxZ}{max}

\begin{document}
\title{A  Quantum Approximate Optimization Algorithm}
\author{Edward Farhi}
\affiliation{Center for Theoretical Physics\\ Massachusetts Institute of Technology\\ Cambridge, MA 02139}
\author{Jeffrey Goldstone}
\affiliation{Center for Theoretical Physics\\ Massachusetts Institute of Technology\\ Cambridge, MA 02139}
\author{Sam Gutmann}

\preprint{MIT-CTP/4610}

\begin{abstract}
We introduce a quantum algorithm that produces approximate solutions for combinatorial optimization problems.  The algorithm depends on an integer $p \geq1$ and the quality of the approximation improves as $p$ is increased. The quantum circuit that implements the algorithm consists of unitary gates whose locality is at most the locality of the objective function whose optimum is sought.  The depth of the circuit grows linearly with $p$ times (at worst) the number of constraints.  If $p$ is fixed, that is, independent of the input size, the algorithm makes use of efficient classical preprocessing.  If $p$ grows with the input size a different strategy is proposed.  We study the algorithm as applied to MaxCut on regular graphs and analyze its performance on 2-regular and 3-regular graphs for fixed $p$.  For $p = 1$, on 3-regular graphs the quantum algorithm always finds a cut that is at least 0.6924 times the size of the optimal cut.  
\end{abstract}

\maketitle

\section{introduction}

Combinatorial optimization problems are specified by $n$ bits and $m$ clauses.  Each clause is a constraint on a subset of the bits which is satisfied for certain assignments of those bits and unsatisfied for the other assignments.  The objective function, defined on $n$ bit strings, is the number of satisfied clauses, 
\begin{equation}
C(z) = \sum\limits^m_{\alpha=1}\, C_\alpha (z)
\label{FarhiEq1}
\end{equation}
where $z= z_1 z_2\ldots z_n$ is the bit string and $C_{\alpha} (z) = 1$ if $z$ satisfies clause $\alpha$ and 0 otherwise.  Typically $C_\alpha$ depends on only a few of the $n$ bits. Satisfiability asks if there is a string that satisfies every clause.   MaxSat asks for a string that maximizes the objective function.   Approximate optimization asks for a string $z$ for which $C(z)$ is close to the maximum of $C$.  In this paper we present a general quantum algorithm for approximate optimization.   We study its performance in special cases of MaxCut and also propose an alternate form of the algorithm geared toward finding a large independent set of vertices of a graph.

The quantum computer works in a $2^n$ dimensional Hilbert space with computational basis vectors $|z\rangle$, and we view \eqref{FarhiEq1} as an operator which is diagonal in the computational basis.  Define a unitary operator $U (C,\gamma)$ which depends on an angle $\gamma$,
\begin{equation}
U (C,\gamma) = e^{-i\gamma C} = \prod\limits^{m}_{\alpha =1}\, e^{-i\gamma C_\alpha } \;.
\label{FarhiEq3}
\end{equation}
All of the terms in this product commute because they are diagonal in the computational basis and each term's locality is the locality of the clause $\alpha$.  Because $C$ has integer eigenvalues we can restrict $\gamma$ to lie between 0 and $2\pi$.  Define the operator $B$ which is the sum of all single bit $\sigma^x $ operators,
\begin{equation}
B = \sum\limits^n_{j=1} \sigma^x_j.
\label{FarhiEq4}
\end{equation}
Now define the $\beta$ dependent product of commuting one bit operators
\begin{equation}
U (B, \beta) = e^{-i\beta B} = \prod\limits^n_{j=1}\, e^{-i \beta  \sigma^x_j}
\label{FarhiEq5}
\end{equation}
where $\beta$ runs from $0$ to $\pi$.  The initial state $ \ket{s} $ will be the uniform superposition over computational basis states:
\begin{equation}
\ket{s} = \frac{1}{\sqrt{2^n}}\ \sum\limits_z \ket{z} .
\label{FarhiEq6}
\end{equation}
For any integer $ p \geq 1$ and $2p$ angles $\gamma_1 \ldots \gamma_p \equiv \boldsymbol{\gamma} \ \text{and}\ \beta_1\ldots \beta_p \equiv \boldsymbol{\beta}$  we define the angle dependent quantum state:
\begin{equation}
\ket{\boldsymbol{\gamma}, \boldsymbol{\beta}} = U (B, \beta_p) \, U (C, \gamma_p) \cdots U (B, \beta_1) \, U (C, \gamma_1)\, \ket{s} .
\label{FarhiEq7}
\end{equation}
Even without taking advantage of the structure of the instance, this state can be produced by a quantum circuit of depth at most $m p + p$. Let $F_p$ be the expectation of $C$ in this state 
\begin{equation}
F_p (\boldsymbol{\gamma}, \boldsymbol{\beta}) = \bra{\boldsymbol{\gamma}, \boldsymbol{\beta}} C  \ket{\boldsymbol{\gamma}, \boldsymbol{\beta}} .
\label{FarhiEq8}
\end{equation}
and let $M_p$ be the maximum of $F_p$ over the angles,
\begin{equation}
M_p = \maxZ_{\boldsymbol{\gamma}, \boldsymbol{\beta}}\, F_p\, (\boldsymbol{\gamma}, \boldsymbol{\beta}).
\label{FarhiEq9}
\end{equation}
Note that the maximization at $p-1$ can be viewed as a constrained maximization at $p$ so 
\begin{equation}
M_p \geq M_{p-1}.
\label{FarhiEq10}
\end{equation}
Furthermore we will later show that
\begin{equation}
\lim_{p \to \infty} M_p = \maxZ_z\, C(z) .
\label{FarhiEq11}
\end{equation}

These results suggest a way to design an algorithm.  Pick a $p$ and start with a set of angles ($\boldsymbol{\gamma}, \boldsymbol{\beta}$) that somehow make $F_p$ as large as possible. Use the quantum computer to get the state $\ket{\boldsymbol{\gamma}, \boldsymbol{\beta}}$.  Measure in the computational basis to get a string $z$ and evaluate $C(z)$. Repeat with the same angles. Enough repetitions will produce a string $z$ with $C(z)$ very near or greater than $F_p (\boldsymbol{\gamma}, \boldsymbol{\beta})$. The rub is that it is not obvious in advance how to pick good angles.  

If $p$ doesn't grow with $n$, one possibility is to run the quantum computer with  angles ($\boldsymbol{\gamma}, \boldsymbol{\beta}$) chosen from a fine grid on the compact set $[0,2\pi]^{p}\times[0,\pi]^p$, moving through the grid to find the maximum of $F_p$. Since the partial derivatives of $F_p (\boldsymbol{\gamma}, \boldsymbol{\beta})$ in \eqref{FarhiEq8} are bounded by $\mathcal{O} (m^2 + m n)$ this search will efficiently produce a string $z$ for which $C(z)$ is close to $M_p$ or larger. However we show in the next section that if $p$ does not grow with $n$ and each bit is involved in no more than a fixed number of clauses, then there is an efficient classical calculation that determines the angles that maximize $F_p$. These angles are then used to run the quantum computer to produce the state $\ket{\boldsymbol{\gamma}, \boldsymbol{\beta}}$ which is measured in the computational basis to get a string $z$. The mean of $C(z)$ for strings obtained in this way is $M_p$.

\section{Fixed $\protect\scalebox{1.25}{$\mathit{p}$}$ Algorithm}
 We now explain how for fixed $p$ we can do classical preprocessing and determine the angles $\boldsymbol{\gamma}$ and $\boldsymbol{\beta}$ that maximize $F_p (\boldsymbol{\gamma}, \boldsymbol{\beta})$. This approach will work more generally but we illustrate it for a specific problem, MaxCut for graphs with bounded degree. The input is a graph with $n$ vertices and an edge set $\left\{\brake{j k}\right\}$ of size $m$. The goal is to find a string $z$ that makes 
\begin{equation}
C = \sum_{\langle j k\rangle} C_{\langle j k\rangle},
\label{FarhiEq12}
\end{equation}
\noindent where
\begin{equation}
C_{\langle j k\rangle} = \frac{1}{2}\left (-\sigma^z_j \sigma^z_k +1 \right),
\end{equation}
as large as possible. Now
\begin{equation}
F_p (\boldsymbol{\gamma}, \boldsymbol{\beta}) = \sum_{\brake{j k}} \bra{s} U^\dagger (C, \gamma_1) \cdots U^\dagger (B, \beta_p)\, C_{\langle j k\rangle} U (B, \beta_p) \cdots U (C, \gamma_1) \ket{s} .
\label{FarhiEq13} 
\end{equation}
Consider the operator associated with edge $\brake{j k }$
\begin{equation}
U^\dagger (C, \gamma_1) \cdots U^\dagger (B, \beta_p) C_{\langle j k\rangle} U (B, \beta_p) \cdots U (C, \gamma_1).
\label{FarhiEq14}
\end{equation}
This operator only involves qubits $j$ and $k$ and those qubits whose distance on the graph from $j$ or $k$ is less than or equal to $p$. To see this consider $p=1$ where the previous expression is
\begin{equation}
U^\dagger (C, \gamma_1)\, U^\dagger (B, \beta_1) C_{\langle j k\rangle} U (B, \beta_1) \, U (C, \gamma_1) .
\label{FarhiEq15}
\end{equation}
The factors in the operator $U (B, \beta_1)$ which do not involve qubits $j$ or $k$ commute through $C_{\langle j k \rangle}$ and we get
\begin{equation}
U^\dagger (C, \gamma_1)\, e^{i \beta_1( \sigma^x_j + \sigma^x_k)}C_{\langle j k\rangle} e^{-i \beta_1 (\sigma^x_j + \sigma^x_k)}\, U (C, \gamma_1) .
\label{FarhiEq16}
\end{equation}
Any factors in the operator  $U (C,\gamma_1)$ which do not involve qubits $j$ or $k$  will commute through and cancel out. So the operator in equation \eqref{FarhiEq16} only involves 
the edge $\brake{jk}$ and edges adjacent to $\brake{jk}$, and qubits on those edges. For any $p$ we see that the operator in \eqref{FarhiEq14} only involves edges at most $p$ steps away from $\brake{jk}$ and qubits on those edges.

Return to equation \eqref{FarhiEq13} and note that the state $\ket{s}$  is the product of $\sigma^x$ eigenstates
\begin{equation}
\ket{s} = \ket{+}_1 \ket{+}_2 \ldots \ket{+}_n
\label{FarhiEq17}
\end{equation}
so each term in equation \eqref{FarhiEq13} depends only on the subgraph involving qubits $j$ and $k$ and those at a distance no more than $p$ away.  These subgraphs each contain a number of qubits that is independent of $n$ (because the degree is bounded) and this allows us to evaluate $F_p$ in terms of quantum subsystems whose sizes are independent  of $n$.  

As an illustration consider MaxCut restricted to input graphs of  fixed degree 3.  For $p=1$, there are only these possible subgraphs for the edge $\brake{jk}$:

\begin{equation}
\includegraphics[scale=.95,trim=50 0 0 0]{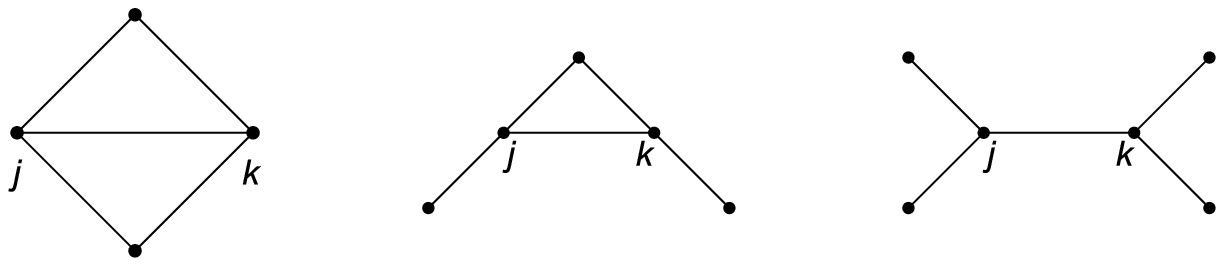}\label{eq:ggg}
\end{equation}

\noindent We will return to this case later.  

For any subgraph $G$ define the operator $C_G$ which is $C$ restricted to $G$,
\begin{equation}
C_G = \sum\limits_{\brake{\ell \ell^\prime}  \epsilon G}C_{\langle \ell \ell^\prime\rangle},
\label{FarhiEq19}
\end{equation}
and the associated operator 
\begin{equation}
U (C_G, \gamma) = e^{-i\gamma \, C_G} .
\label{FarhiEq20}
\end{equation}
Also define
\begin{equation}
B_G = \sum\limits_{j \epsilon G} \sigma^x_j
\label{FarhiEq21}
\end{equation}
and
\begin{equation}
U (B_G, \beta) = e^{-i\beta B_G} .
\label{FarhiEq22}
\end{equation}
Let the state $\ket{s, G}$ be
\[
\ket{s, G} = \prod\limits_{\ell \epsilon G} \ \ket{+}_\ell .
\]
Return to equation \eqref{FarhiEq13}. Each edge $\brake{j, k}$ in the sum is associated with a subgraph $g (j, k)$ and makes a contribution to $F_p$ of 
\begin{multline}
\bra{s, g (j, k)} U^\dagger \, (C_{g (j, k)}, \gamma_p) \cdots U^\dagger (B_{g (j, k)}, \beta_1)C_{\langle j k\rangle}U \, (B_{g (j, k)}, \beta_1) \cdots U\, (C_{g (j, k)}, \gamma_p)
	\ket{s, g (j, k)}
		\label{FarhiEq23}
\end{multline}
 The sum in \eqref{FarhiEq13} is over all edges, but if two edges $\brake{j k}$ and $\brake{j^\prime k^\prime}$ give rise to isomorphic subgraphs, then the corresponding functions of ($\boldsymbol{\gamma}, \boldsymbol{\beta}$) are the same. Therefore we can view the sum in \eqref{FarhiEq13} as a sum over subgraph types.  Define
\begin{multline}
f_g\, (\boldsymbol{\gamma}, \boldsymbol{\beta}) = \bra{s, g (j, k)} U^\dagger(C_{g (j,k)}, \gamma_1) \cdots U^\dagger (B_{g (j, k)}, \beta_p)C_{\langle j k\rangle} U(B_{g(j,x)}\beta_p) \cdots\\ U(C_{g (j, k)}, \gamma_1) \ket{s, g (j, k)},
\label{FarhiEq24}
\end{multline}
where $g (j,k)$ is a subgraph of type $g$. $F_p$ is then
\begin{equation}
F_p \, (\boldsymbol{\gamma}, \boldsymbol{\beta}) = \sum\limits_g \, w_g\, f_g (\boldsymbol{\gamma}, \boldsymbol{\beta})
\label{FarhiEq25}
\end{equation}
where $w_g$ is the number of occurrences of the subgraph $g$ in the original edge sum.  The functions $f_g$ do not depend on $n$ and $m$.  The only dependence on $n$ and $m$ comes through the weights $w_g$ and these are just read off the original graph.   Note that the expectation in \eqref{FarhiEq24} only involves the qubits in subgraph type $g$.  The maximum number of qubits that can appear in \eqref{FarhiEq23} comes when the subgraph is a tree.  For a graph with maximum degree $v$, the numbers of qubits in this tree is
\begin{equation}
q_{\text{tree}} = 2 \left[ \frac{(v-1)^{p+1} -1}{(v-1) -1}\right],
\label{FarhiEq26}
\end{equation}
(or $2p + 2 \ \text{if}\ v=2$), which is $n$ and $m$ independent. For each $p$ there are only finitely many subgraph types.

Using \eqref{FarhiEq24}, $F_p(\boldsymbol{\gamma}, \boldsymbol{\beta})$ in \eqref{FarhiEq25} can be evaluated on a classical computer whose  resources are not growing with $n$.  Each $f_g$ involves operators and states in a Hilbert space whose dimension is at most $2^{q_{\text{tree}}}$.  Admittedly for large $p$ this may be beyond current classical technology, but the resource requirements do not grow with $n$. 

To run the quantum algorithm we  first find the ($\boldsymbol{\gamma},\boldsymbol{\beta}$) that maximize $F_p$.  The only dependence on $n$ and $m$ is in the weights $w_g$ and these are easily evaluated.  Given the best ($\boldsymbol{\gamma}, \boldsymbol{\beta}$) we turn to the quantum computer and produce the state $\ket{\boldsymbol{\gamma}, \boldsymbol{\beta}}$  given in equation \eqref{FarhiEq7}.   We then measure in the computational basis and get a string $z$ and evaluate $C(z)$.  Repeating gives a sample of values of $C(z)$ between $0$ and $+m$ whose mean is $F_p (\boldsymbol{\gamma}, \boldsymbol{\beta})$. An outcome of at least $F_p (\boldsymbol{\gamma}, \boldsymbol{\beta}) - 1$ will be obtained with probability $1-1/m$ with order $m \log m$ repetitions.  

\section{Concentration}
Still using MaxCut on regular graphs as our example, it is useful to get information about the spread of $C$ measured in the state $\ket{\boldsymbol{\gamma}, \boldsymbol{\beta}}$.
If $v$ is fixed and $p$ is fixed (or grows \mbox{slowly with $n$)} the distribution of $C(z)$ is actually concentrated near its mean. To see this, calculate 
\begin{equation}
\bra{\boldsymbol{\gamma}, \boldsymbol{\beta}} C^2\ket{\boldsymbol{\gamma}, \boldsymbol{\beta}} - \bra{\boldsymbol{\gamma}, \boldsymbol{\beta}} C \ket{\boldsymbol{\gamma}, \boldsymbol{\beta}}^2
\label{FarhiEq27}
\end{equation}
\begin{multline}
= \sum\limits_{\substack{\brake{jk}\\ \brake{j^\prime k^\prime}}} \Bigg[  \bra{s} U^\dagger (C, \gamma_1) \cdots U^\dagger (B, \beta_p)\,C_{\langle j k\rangle}\,C_{\langle j^\prime k^\prime\rangle}\, U (B, \beta_p) \cdots U (C, \gamma_1) \ket{s}\\[-1.5ex]
	- \bra{s} U^\dagger (C, \gamma_1) \cdots U^\dagger (B, \beta_p)\, C_{\langle j k\rangle}\, U (B, \beta_p) \cdots U (C, \gamma_1) \ket{s} \\
	\cdot \bra{s} U^\dagger (C, \gamma_1) \cdots U^\dagger (B, \beta_p)\, C_{\langle j^\prime k^\prime\rangle}\, U (B, \beta_p) \cdots U (C, \gamma_1) \ket{s} \Bigg] .
\label{FarhiEq29}
\end{multline}
If the subgraphs $g(j,k)$ and $g(j^\prime,k^\prime)$ do not involve any common qubits, the summand in \eqref{FarhiEq29} will be 0. The subgraphs $g(j,k)$ and $g(j^\prime,k^\prime)$ will have no common qubits as long as there is no path in the instance graph  from $\brake{jk}$ to $\brake{j^\prime k^\prime}$ of length $2p + 1$ or shorter. From \eqref{FarhiEq26} with $p$ replaced by $2p+1$ we see that for each $\brake{jk}$ there are at most
\begin{equation}
 2 \left[ \frac{(v-1)^{2p+2} -1}{(v-1) -1}\right]
\label{FarhiEq30}
\end{equation}
edges $\brake{j^\prime k^\prime}$ which could contribute to the sum in \eqref{FarhiEq29} (or $4p+4$ if $v=2$) and therefore
\begin{equation}
\bra{\boldsymbol{\gamma}, \boldsymbol{\beta}} C^2\ket{\boldsymbol{\gamma}, \boldsymbol{\beta}} - \bra{\boldsymbol{\gamma}, \boldsymbol{\beta}} C \ket{\boldsymbol{\gamma}, \boldsymbol{\beta}}^2 \leqslant 2 \left[ \frac{(v-1)^{2p+2} -1}{(v-1) -1}\right] \cdot m
\label{FarhiEq31}
\end{equation}
since each summand is at most $1$ in norm. For $v$ and $p$ fixed we see that the standard deviation of $C(z)$ is at most of order $\sqrt{m}$. This implies that the sample mean of order $m^2$ values of $C(z)$ will be within 1 of $F_p (\boldsymbol{\gamma}, \boldsymbol{\beta})$ with probability $1- \frac{1}{m}$. The concentration of the distribution of $C(z)$ also means that there is only a small probability that the algorithm will produce strings with $C(z)$ much bigger than $F_p (\boldsymbol{\gamma}, \boldsymbol{\beta})$.

\section{The Ring of Disagrees}

	We now analyze the performance of the quantum algorithm for MaxCut on 2-regular graphs.  Regular of degree 2 (and connected) means that the graph is a ring.  The objective operator is again given by equation \eqref{FarhiEq12} and its maximum is $n$ or $n-1$ depending on whether $n$ is even or odd.  We will analyze the algorithm for all $p$.

	For any $p$ (less than $n/2$), for each edge in the ring, the subgraph of vertices within $p$ of the edge is a segment of $2p+2$ connected vertices with the given edge in the middle. So for each $p$ there is only one type of subgraph, a line segment of $2p+2$ qubits and the weight for this subgraph type is $n$. We numerically maximize the function given in \eqref{FarhiEq24} and we find that for $p=1, 2 ,3 ,4 ,5$ and 6 the maxima are 3/4, 5/6, 7/8, 9/10, 11/12, and 13/14 to 13 decimal places from which we conclude that $M_p=n(2p+1)/(2p+2)$ for all $p$.  So the quantum algorithm will find a cut of size  $n (2p+1)/(2p+2) - 1$ or bigger.  Since the best cut is $n$, we see that our quantum algorithm can produce an approximation ratio that can be made arbitrarily close to 1 by making $p$ large enough, independent of $n$.  For each $p$ the circuit depth can be made $3p$ by breaking the edge sum in $C$ into two sums over $\brake{ j, j+1}$ with $j$ even and $j$ odd. So this algorithm has a circuit depth independent of $n$.

\section{MaxCut on 3-Regular Graphs}\label{sec:maxcut}
 
We now look at how the Quantum Approximate Optimization Algorithm, the QAOA, performs on MaxCut on (connected) 3-regular graphs.  The approximation ratio is $C(z)$, where $z$ is the output of the quantum algorithm, divided by the maximum of $C$.  We first show that for $p=1$, the worst case approximation ratio that the quantum algorithm produces is $0.6924$. 

 Suppose a 3-regular graph with $n$ vertices (and accordingly $3n/2$ edges) contains $T$ ``isolated triangles'' and $S$ ``crossed squares'', which are subgraphs of the form,

\begin{equation}
\includegraphics[scale=1]{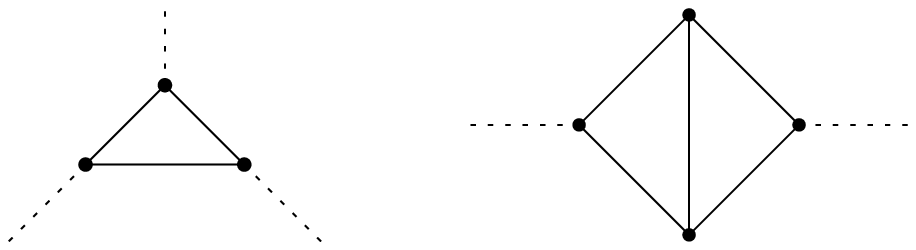}.
\end{equation}

\noindent The dotted lines indicate edges that leave the isolated triangle and the crossed square.   To say that the triangle is isolated is to say that the 3 edges that leave the triangle end on distinct vertices.  If the two edges that leave the crossed square are in fact the same edge, then we have a 4 vertex disconnected  3-regular graph. For this special case (the only case where the analysis below does not apply) the approximation ratio is actually higher than $0.6924$.  In general, $3 T+4 S \leq n$ because no isolated triangle and crossed square can share a vertex.

Return to the edge sum in $F_1(\gamma,\beta)$ of equation \eqref{FarhiEq13}.  For each crossed square there is one edge $\langle j k \rangle$  for which $g(j,k)$ is the first type displayed in \eqref{eq:ggg}.  Call this subgraph type $g_4$ because it has 4 vertices.  In each crossed square there are 4 edges that give rise to subgraphs of the second type displayed in \eqref{eq:ggg}.  We call this subgraph type $g_5$ because it has 5 vertices.  All 3 of the edges in any isolated triangle have subgraph type $g_5$, so there are $4S+3T$ edges with subgraph type $g_5$.  The remaining edges in the graph all have a subgraph type like the third one displayed in \eqref{eq:ggg} and we call this subgraph type $g_6$.  There are $(3n/2 - 5S - 3T)$ of these so we have
\begin{equation}\label{eq:farhi34}
F_1(\gamma,\beta) = S f_{g_4}(\gamma,\beta) + \left(4 S + 3 T\right)f_{g_5}(\gamma,\beta) + \left(\frac{3n}{2} - 5S - 3T \right)f_{g_6}(\gamma,\beta) 
\end{equation}
\noindent The maximum of $F_1$ is a function of $n$, $S$, and $T$,
\begin{equation}
M_1(n,S,T) = \max_{\gamma,\beta} F_1(\gamma,\beta).
\end{equation}
Given any 3 regular graph it is straightforward to count $S$ and $T$.  Then using a classical computer it is straightforward to calculate $M_1(n,S,T)$.  Running a quantum computer with the maximizing angles $\gamma$ and $\beta$ will produce the state $|\gamma, \beta \rangle$ which is then measured in the computational basis.   With order $n \log n$ repetitions a string will be found whose cut value is very near or larger than $M_1(n,S,T)$.  

To get the approximation ratio we need to know the best cut that can be obtained for the input graph.  This is not just a function of $S$ and $T$.  However a graph with $S$ crossed squares and $T$ isolated triangles must have at least one unsatisfied edge per crossed square and one unsatisfied edge per isolated triangle so the number of satisfied edges is $\leq (3n/2 - S - T)$.    This means that for any graph characterized by $n$, $S$ and $T$ the quantum algorithm will produce an approximation ratio that is at least 
\begin{equation}
\frac{M_1(n,S,T)}{\left( \frac{3n}{2} - S - T \right)} \label{eq:Y5}.
\end{equation}
It is convenient to scale out $n$ from the top and bottom of \eqref{eq:Y5}.  Note that $M_1/n$ which comes from $F_1/n$ depends only on $S/n \equiv s$ and $T/n \equiv t$.  So we can write \eqref{eq:Y5} as
\begin{equation}
\frac{M_1(1,s,t)}{\left( \frac{3}{2} - s - t \right)} \label{eq:Y6}
\end{equation}
where $s,t \geq 0$ and $4s + 3t \leq 1.$  It is straightforward to numerically evaluate \eqref{eq:Y6} and we find that it achieves its minimum value at $s = t = 0$ and the value is $0.6924$.  So we know that on any 3-regular graph, the QAOA will always produce a cut whose size is at least $0.6924$ times the size of the optimal cut.  This $p=1$ result on 3-regular graphs is not as good as known classical algorithms \cite{halperin-2004}.

It is possible to analyze the performance of the QAOA for $p = 2$ on 3-regular graphs.  However it is more complicated then the $p = 1$ case and we will just show partial results.  The subgraph type with the most qubits is this tree with 14 vertices:

\begin{align}
\includegraphics[scale=0.5,trim=30 0 0 0]{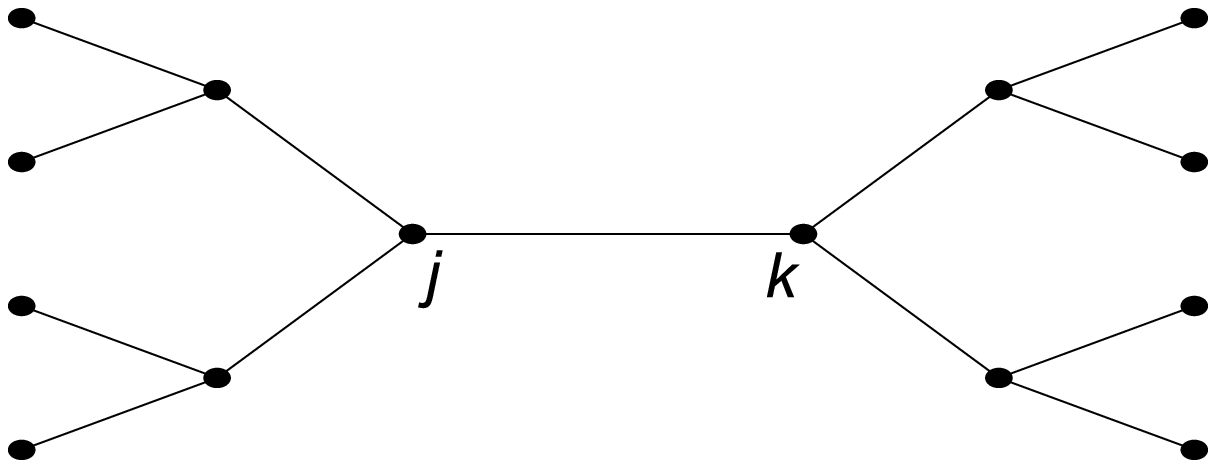} \label{g14}
\end{align}

Numerically maximizing \eqref{FarhiEq24} with $g$ given by \eqref{g14} yields $0.7559$. Consider a 3-regular graph on $n$ vertices with $o(n)$  pentagons, squares and triangles.   Then all but $o(n)$ edges have \eqref{g14} as their subgraph type.  The QAOA at $p=2$ cannot detect whether the graph is bipartite, that is, completely satisfiable, or contains many odd loops of length 7 or longer.  If the graph is bipartite the approximation ratio is $0.7559$ in the limit of large $n$.  If the graph contains many odd loops (length $7$ or more), the approximation ratio will be higher.

\section{Relation to the Quantum Adiabatic Algorithm}\label{sec:QAA}

We are focused on finding a good approximate solution to an optimization problem whereas the Quantum Adiabatic Algorithm, QAA \cite{farhi-2000},  is designed to find the optimal solution and will do so if the run time is long enough. Consider the time dependent Hamiltonian $H(t)=(1-t/T) B+(t/T) C$.  Note that the state $\ket{s}$  is the highest energy eigenstate of $B$ and we are seeking a high energy eigenstate of $C$.  Starting in $\ket{s}$ we could run the quantum adiabatic algorithm and if the run time $T$ were long enough we would find the highest energy eigenstate of $C$.  Because $B$ has only non-negative off-diagonal elements, the Perron-Frobenius theorem implies that the difference in energies between the top state and the one below is greater than 0 for all $t < T$, so for sufficiently large $T$ success is assured.   A Trotterized approximation to the evolution consists of an alternation of the operators $U(C,\gamma)$ and $U(B,\beta)$ where the sum of the angles is the total run time.  For a good approximation we want each $\gamma$ and $\beta$ to be small and for success we want a long run time so together these force $p$ to be large. In other words, we can always find a $p$ and a set of angles $\boldsymbol{\gamma}, \boldsymbol{\beta}$ that make $F_p(\boldsymbol{\gamma}, \boldsymbol{\beta})$ as close to $M_p$ as desired.   With \eqref{FarhiEq10}, this proves the assertion of \eqref{FarhiEq11}.  

The previous discussion shows that we can get a good approximate solution to an optimization problem by making $p$ sufficiently large, perhaps exponentially large in $n$.  But the QAA works by producing a state with a large overlap with the optimal string. In this sense \eqref{FarhiEq11}, although  correct, may be misleading.  In fact on the ring of disagrees the state produced at $p=1$, which gives a 3/4 approximation ratio, has an exponentially small overlap with the optimal strings.  

We also know an example where the QAA fails and the QAOA succeeds. In this example (actually a minimization) the objective function is symmetric in the $n$ bits and therefore depends only on the Hamming weight. The objective function is plotted in figure 1 of reference \cite{farhi-2002}. Since the beginning Hamiltonian is also symmetric the evolution takes place in a subspace of dimension $n+1$ with a basis of states $|w\rangle$ indexed by the Hamming weight. The example can be simulated and analyzed for large $n$. For subexponential run times, the QAA is trapped in a false minimum at $w=n$. The QAOA can be similarly simulated and analyzed.  For large $n$, even with $p=1$, there are values of $\gamma_1$ and $\beta_1$ such that the final state is concentrated near the true minimum at $w=0$.

The Quantum Approximate Optimization Algorithm has the key feature that as $p$ increases the approximation improves.  We contrast this to the performance of the QAA.  For realizations of the QAA there is a total run time $T$ that also appears in the instantaneous Hamiltonian,  $H(t)=\widetilde{H} (t/T)$.  We start in the ground state of $\widetilde{H} (0)$ seeking the ground state of $\widetilde{H}  (1)$.  As $T$ goes to infinity the overlap of the evolved state with the desired state goes to 1.  However the success probability is generally not a monotonic function of $T$.  See figure 2 of reference \cite{crosson-2014} for an extreme example where the success probability is plotted as a function of $T$ for a particular 20 qubit instance of Max2Sat. The probability rises and then drops dramatically,  and the ultimate rise for large $T$ is not seen for times that can be reasonably simulated. It may well be advantageous in designing strategies for the QAOA to use the fact that the approximation improves as $p$ increases.

\section{A Variant of the Algorithm}
We are now going to give a variant of the basic algorithm which is suited to situations where the search space is a complicated subset of the $n$ bit strings.  We work with an example that illustrates the basic idea.  Consider the problem of finding a large independent set in a given graph of $n$ vertices.  An independent set is a subset of the vertices with the property that no two vertices in the subset have an edge between them.  With the vertices labeled 1 to $n$, a subset of the vertices corresponds to the string $z= z_1z_2 \ldots z_n$ with each bit being 1 if the corresponding vertex is in the subset and the bit is 0 if the vertex is not.  We restrict to strings which correspond to independent sets in the graph.   The size of the independent set is the Hamming weight of the string $z$ which we denote by $C(z)$,

\begin{equation}
C(z) = \sum_{j = 1}^n z_j,
\end{equation}

\noindent and the goal is to find a string $z$ that makes $C(z)$ large.

The Hilbert space for our quantum algorithm has an orthonormal basis $| z \rangle$ where $z$ is any string corresponding to an independent set.  In cases of interest, the Hilbert space dimension is exponentially large in $n$, though not as big as $2^n$.  The Hilbert space is not a simple tensor product of qubits.  The operator $C$ is associated with the $\gamma$ dependent unitary
\begin{equation}
U(C,\gamma) = e^{-i \gamma C}
\end{equation}
where $\gamma$ lies between 0 and $2\pi$ because $C$ has integer eigenvalues.  We define the quantum operator $B$ that connects the basis states:
\begin{equation}\label{eq:adjacency}
\langle z | B | z' \rangle = \left\{
     \begin{array}{lr}
       1 :  z \textnormal{ and }z' \textnormal{ differ in one bit}\\
       0 : \textnormal{otherwise\indent\indent\indent\indent\indent\indent\indent.}
     \end{array}
   \right. 
\end{equation}
Note that $B$ is the adjacency matrix of the hypercube restricted to the legal strings, that is, those that correspond to independent sets in the given graph.  Now, in general, $B$ does not have integer eigenvalues so we define
\begin{equation}
U(B,b) = e^{-i b B}
\end{equation}
\noindent where $b$ is a real number.  

For the starting state of our algorithm we take the easy to construct state $|z\!=\!0\rangle$  corresponding to the empty independent set which has the minimum value of $C$.  For $p \geq 1$, we have $p$ real numbers $b_1, b_2 \ldots  b_p \equiv \mathbf{b}$ and $p-1$ angles $\gamma_1, \gamma_2, ….\gamma_{p-1} \equiv \boldsymbol{\gamma}$.  The quantum state 
\begin{equation}
|\mathbf{b},\boldsymbol{\gamma}\rangle = U(B,b_p)U(C,\gamma_{p-1})\cdots U(B,b_1)|z\!=\!0\rangle \label{eq:lotsu}
\end{equation}
\noindent is what we get after the application of an alternation of the operators associated with $B$ and $C$.  Now we can define
\begin{equation}
F_p(\mathbf{b},\boldsymbol{\gamma}) = \langle \mathbf{b},\boldsymbol{\gamma}|C|\mathbf{b},\boldsymbol{\gamma}\rangle \label{eq:fexpect}
\end{equation}
\noindent as the expectation of $C$ in the state  $|\mathbf{b}, \boldsymbol{\gamma}\rangle$.    And finally we define the maximum,
\begin{equation}
M_p = \max_{\mathbf{b},\boldsymbol{\gamma}} F_p(\mathbf{b},\boldsymbol{\gamma}). \label{eq:fmax}
\end{equation}
The maximization at $p-1$ is the maximization at $p$ with $b_p=0$ and $\gamma_{p-1} = 0$ so we have
\begin{equation}
M_p \geq M_{p-1}.
\end{equation}
\noindent Furthermore,
\begin{equation}
\lim_{p \rightarrow \infty} M_p = \max_{z{\small \textnormal{ legal}}} C(z). \label{eq:lim}
\end{equation}

To see why \eqref{eq:lim} is true note that the initial state is the ground state of $C$, which we view as the state with the maximum eigenvalue of $-C$.   We are trying to reach a state which is an eigenstate of $+C$ with maximum eigenvalue.  There is an adiabatic path (which stays at the top of the spectrum throughout) with run time $T$ that achieves this as $T$ goes to infinity.   This path has two parts.  In the first we interpolate between the beginning Hamiltonian $-C$ and the Hamiltonian $B$,
\begin{equation}
H(t) = \left(1- \frac{2t}{T}\right)(-C) + \frac{2t}{T}B \; \; ,  \; \; \; \; 0 \leq t \leq \frac{T}{2}
\end{equation}
\noindent We evolve the initial state with this Hamiltonian for time $T/2$ ending arbitrarily close to the top state of $B$.   Next we interpolate between the Hamiltonian $B$ and the Hamiltonian $+C$,
\begin{equation}
H(t) = \left(2 - \frac{2t}{T}\right)B + \left(\frac{2t}{T}-1\right)C\; \; ,  \; \; \; \;  \frac{T}{2} \leq t \leq T
\end{equation}
\noindent evolving the quantum state just produced from time $t=T/2$ to $t=T$.  As in section \ref{sec:QAA}, using the Perron-Frobenius Theorem, the Adiabatic Theorem and Trotterization we get the result given in \eqref{eq:lim}.

Together \eqref{eq:lotsu} through \eqref{eq:lim} suggest a quantum subroutine for an independent set algorithm.  For a given $p$ and a given $(\mathbf{b},\boldsymbol{\gamma})$ produce the quantum state  $| \mathbf{b} , \boldsymbol{\gamma} \rangle$  of \eqref{eq:lotsu}.   Measure in the computational basis to get a string $z$ which labels an independent set whose size is the Hamming weight of $z$.  Repeat with the same $(\mathbf{b} , \boldsymbol{\gamma})$ to get an estimate of $F_p (\mathbf{b},\boldsymbol{\gamma})$ in \eqref{eq:fexpect}.  This subroutine can be called by a program whose goal is to get close to $M_p$ given by \eqref{eq:fmax}.   This enveloping program can be designed using either the methods outlined in this paper or novel techniques.

For $p=1$, the subroutine can be thought of as evolving the initial state $|z\!=\!0\rangle$ with the Hamlitonian $B$ for a time $b$.  $B$ is the adjacency matrix of a big graph whose vertices correspond to the independent sets of the input graph and whose edges can be read off \eqref{eq:adjacency}.  We view this as a continuous time quantum walk entering the big graph at a particular vertex \cite{farhi-2007}. In the extreme case where the input graph has no edges, all strings of length $n$ represent independent sets so the Hilbert space dimension is $2^n$.  In this case $B$ is the adjacency matrix of the hypercube, realizable as in \eqref{FarhiEq4}.  Setting $b=\pi / 2$, the state \eqref{eq:lotsu} (with $p=1$ there is only one unitary) is  $|z=11\ldots 11\rangle$  which maximizes the objective function.   In the more general case we can view \eqref{eq:lotsu} as a succession of quantum walks punctuated by applications of $C$ dependent unitaries which aid the walk in achieving its objective.  The algorithm of the previous sections can also be viewed this way although the starting state is not a single vertex.  

\section{Conclusion}
We introduced a quantum algorithm for approximate combinatorial optimization that depends on an integer parameter $p$. The input is an $n$ bit instance with an objective function $C$ that is the sum of $m$ local terms.  The goal is to find a string $z$ for which $C(z)$ is close to $C$'s global maximum.  In the basic algorithm, each call to the quantum computer uses a set of $2p$ angles $(\boldsymbol{\gamma}, \boldsymbol{\beta})$ and produces the state 
\begin{equation}
\ket{\boldsymbol{\gamma}, \boldsymbol{\beta}} = U (B, \beta_p) \, U (C, \gamma_p) \cdots U (B, \beta_1) \, U (C, \gamma_1)\, \ket{s} .
\label{FarhiEq32}
\end{equation}
This is followed by a measurement in the computational basis yielding a string $z$ with an associated value $C(z)$.  Repeated calls to the quantum computer will yield a good estimate of 
\begin{equation}
F_p (\boldsymbol{\gamma}, \boldsymbol{\beta}) = \bra{\boldsymbol{\gamma}, \boldsymbol{\beta}} C  \ket{\boldsymbol{\gamma}, \boldsymbol{\beta}} .
\label{FarhiEq33}
\end{equation}
Running the algorithm requires a strategy for a picking a sequence of sets of angles with the goal of making $F_p$ as big as possible. We give several possible strategies for finding a good set of angles.

In section II we focused on fixed $p$ and the case where each bit is in no more than a fixed number of clauses.  In this case there is an efficient classical algorithm that determines the best set of angles which is then fed to the quantum computer.  Here the quantum computer is run with only the best set of angles. Note that the ``efficient" classical algorithm which  evaluates \eqref{FarhiEq25} using \eqref{FarhiEq24} could require space doubly exponential in $p$.

An alternative to using a classical preprocessor to find the best angles is to make repeated calls to the quantum computer with different sets of angles.  One strategy, when $p$ does not grow with $n$ is to put a fine grid on the compact set $[0,2\pi]^p \times [0 , \pi ] ^ {p}$ where the number of points is only polynomial in $n$ and $m$.  This works because the function $F_p$ does not have peaks that are so narrow that they are not seen by the grid.

The QAOA can be run on a quantum computer with $p$ growing with $n$ as long as there is a strategy for choosing sets of angles.  Perhaps for some combinatorial optimization problem, good angles can be discovered in advance.  Or the quantum computer can be called to evaluate $F_p (\boldsymbol{\gamma}, \boldsymbol{\beta})$, the expectation of $C$ in the state $\ket{\boldsymbol{\gamma}, \boldsymbol{\beta}}$.  This call can be used as a subroutine by a classical algorithm that seeks the maximum of the smooth function $F_p(\boldsymbol{\gamma}, \boldsymbol{\beta})$.  We hope that either $p$ fixed or growing slowly with $n$ will be enough to have this quantum algorithm be of use in finding solutions to combinatorial search problems beyond what classical algorithms can achieve.  
\section{Acknowledgements}
This work was supported by the US Army Research Laboratory's Army Research Office
through grant number W911NF-12-1-0486, and the National Science Foundation through grant number
CCF-121-8176.  The authors thank Elizabeth Crosson for discussion and help in preparing the manuscript.  We also thank Cedric Lin and Han-Hsuan Lin for their help.  EF would like to thank the Google Quantum Artificial Intelligence Lab for discussion and support.

\end{document}